# Profilometry with digital fringe-projection at the spatial and temporal Nyquist frequencies

**MOISES PADILLA,**[*] **MANUEL SERVIN AND GUILLERMO GARNICA**

*Centro de Investigaciones en Optica AC, Loma del Bosque 115, Leon, Guanajuato, Mexico.*
[*]*moises@cio.mx*

**Abstract:** A phase-demodulation method for digital fringe-projection profilometry using the spatial and temporal Nyquist frequencies is presented. It allows us to digitize tridimensional surfaces using the highest spatial frequency (π radians per pixel) and consequently with the highest sensitivity for a given digital fringe projector. Working with the highest temporal frequency (π radians per temporal sample), the proposed method rejects the DC component and all even-order distorting harmonics using a bare-minimum 2-step phase shift. The proposed method is suitable for digitization of piece-wise continuous surfaces because it does not require spatial low-pass filtering. Gamma calibration is unnecessary because the projected fringes are binary, and the harmonics produced by the binary profile can be easily attenuated with a slight defocusing on the digital projector. Viability of the proposed method is supported by experimental results showing complete agreement with the predicted behavior.

## 1. Introduction

Nowadays, structured-light projection profilometry is arguably the most popular technique for tri-dimensional (3D) shape measurement [1-3]. Particularly the phase-demodulation methods working with sinusoidal fringe patterns provide the highest-quality dense estimations of the digitized surface [1-4]. A typical setup for fringe-projection profilometry (FPP) requires an image projector, a camera, and a computing device for control and data processing. The sensitivity in FPP is proportional to the spatial frequency of the projected fringes, and to the tangent of the angle between the projection and observation directions [5]. However, using a large angle is generally unadvisable because extensive self-occluding shadows are produced over the 3D solid under study. Likewise, the highest spatial frequency we can work with is limited by the discretization of the projected fringes, and by the lateral-spatial resolution of the camera and the projector. Also, we must remark, modern digital cameras have much higher spatial-resolution in comparison to digital image projectors. Therefore, being able to work with the highest spatial frequency achievable by the digital projector is of great interest.

According to the Nyquist-Shannon sampling theorem, the maximum frequency of a sampled signal must be strictly lower than half the sampling frequency in order to avoid information loss due to aliasing [6]. Half the sampling frequency of a discrete system is called the Nyquist frequency. Digital projector and cameras have a Nyquist frequency 0.5 samples per pixel; thus, a spatial period longer than 2 pixels is necessary when projecting/recording sinusoidal fringes in order to avoid aliasing. Nevertheless, as we will show in this paper, we can work at the spatial Nyquist frequency for the fringe-projection stage (spatial frequency for the recording stage must remain below Nyquist). Describing each spatial period with two discrete samples, the fringes become binary and an analog low-pass filter (defocused projection) must be applied in order to recover a sinusoidal profile [7].

Defocused projection of binary fringes removes the need to calibrate non-linear gamma encoding of most projectors, which is one of the main difficulties for high-quality phase estimations in digital FPP [1-8]. It also impedes us from using many-step phase-shifting algorithms (PSAs) for robust quadrature filtering. Nevertheless, as we will show in this paper, working with the spatial and temporal Nyquist frequencies we derived a phase-demodulation algorithm with reasonable high signal-to-noise ratio (SNR) and as robust against distorting harmonics as the popular 4-step least-squares phase-shifting algorithm (LS-PSA).

The rest of the paper is divided as follows. Section 2 presents the theoretical framework for the proposed method. In section 3 we derive the figures of merit for our proposed method: SNR and robustness against distorting harmonics using the frequency transfer function (FTF) formalism. Section 4 shows some experimental results consistent with the predicted behavior. Finally, section 5 contains our conclusions and some perspectives for future work.

## 2. Theoretical framework

Consider the single-projector single-camera setup for digital fringe-projection profilometry illustrated in Fig. 1. Our phase-shifted open fringes to be projected are modeled as

$$f(x_p, y_p, t) = 0.5 + 0.5\cos(u_0 x_p + \omega_0 t), \tag{1}$$

where $\{x_p, y_p, t\}$ are defined in the integer domain. The spatial and temporal frequencies are respectively $u_0 \in (0, \pi]$ (in radians/pixel) and $\omega_0 \in (0, \pi]$ (in radians per temporal sample). The normalized intensity is assumed to be discretized on at least $2^8$ levels for negligible discretization distortion. Finally, the label "p" stands for projection.

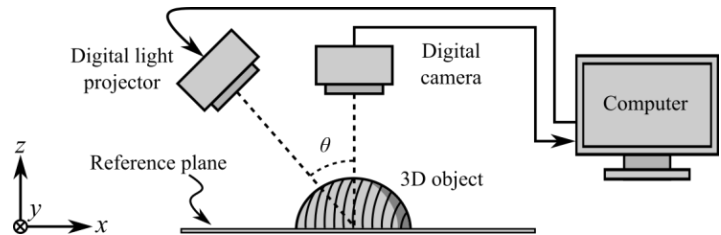

Fig. 1. Basic setup for a single-projector single-camera digital fringe-projector profilometer.

When projected over a 3D object, the phase-shifted open fringes in Eq. (1) become phase-modulated proportionally to the height profile of the surface under test [1-5]. Our ideal mathematical model for the sinusoidal phase-modulated fringe pattern is given by:

$$I(x,y,t) = a(x,y) + b(x,y)\cos[\varphi(x,y) + u'x + \omega_0 t], \quad (x,y) \in \mathbb{Z}^2, \quad t \in \mathbb{Z}. \quad (2)$$

$a(x,y)$ and $b(x,y)$ are the background signal and fringes' contrast functions, respectively. $\varphi(x,y) = u_0 \tan(\theta) s(x,y)$ is the searched modulating phase proportional to the 3D surface $s(x,y)$; and $\theta \in (0, \pi/2)$ is the sensitivity angle. The effective spatial-carrier frequency is $u' = \alpha u_0$, with $0 < \alpha < 1$ because modern cameras have higher pixel resolution in comparison to the digital projectors. For identical spatial resolutions, $\alpha < 1$ remains true if the camera sees just a portion of the region illuminated by the digital projector [9].

Rewriting $I(x,y,t)$ using Euler's formula, we have:

$$I(x,y,t) = a(x,y) + c(x,y)e^{i(u'x+\omega_0 t)} + c^*(x,y)e^{-i(u'x+\omega_0 t)},$$
$$c(x,y) = (1/2)b(x,y)\exp[i\varphi(x,y)]. \quad (3)$$

Both spatial and temporal carriers are assumed to be known ($u'$ is easily calibrated using a reference plane), so we can estimate $c(x,y)$ or $c^*(x,y)$ using synchronous methods. By convention we choose $c(x,y)$, from where the searched phase-map and the fringes' contrast function are computed as its argument and amplitude, respectively [1-5]. For the spatial synchronous method, having $u_0 \in (0, \pi)$, the searched analytic signal is given by

$$r_1(x,y)\exp[i\varphi_1(x,y)] = e^{-i\omega_0 t}[I(x,y,t)e^{-iu'x}] * h(x,y), \quad (4)$$

where $r_1(x,y) \approx b(x,y)$, and $h(x,y)$ is real-valued impulse response with a narrow-bandwidth FTF that isolates the signal of interest around the spectral origin [1-5]. This approach is suitable for dynamic measurements, but the spatial low-pass filtering implies the demodulated phase $\varphi_1(x,y)$ will be distorted around any discontinuity on the surface. Alternatively, for the temporal synchronous method, the searched analytic signal is given by

$$r_2(x,y)\exp[i\varphi_2(x,y)] = e^{-iu'x}[I(x,y,t) * h(t)]_{t=M} = e^{-iu'x}\sum_{m=0}^{M-1} w_m I(x,y,m). \quad (5)$$

Here $r_2(x,y) \propto b(x,y)$ and $M$ represents the number of temporal phase-shifted samples. The complex-valued weight factors $w_n$ define the impulse response function and its FTF as $h(t) = \Sigma w_m \delta(t-m)$, and $H(\omega) = F\{h(t)\} = \Sigma w_m \exp(-im\omega)$. For this phase-demodulation method, $H(\omega)$ is required to fulfill the quadrature conditions [4]:

$$H(-\omega_0) = H(0) = 0, \quad H(\omega_0) \neq 0, \quad \forall \omega_0 \in (0, \pi). \quad (6)$$

Since no spatial low-pass filter is involved, $\varphi_2(x,y)$ is undistorted around discontinuities on the surface. However, requiring several phase-shifted fringe patterns makes this approach unsuitable (in principle) for the study of fast-changing phenomena [1-5].

*2.1 Aliasing at the Nyquist frequencies*

For $u_0 \in (0, \pi)$, excluding the limit values, Eq. (1) describes a sequence of fixed-contrast vertical fringes traveling along the $x_p$-axis. On the other hand, using $u_0 = \pi$ we have

$$\cos(\pi x_p + \omega_0 t) = (-1)^{x_p} \cos(\omega_0 t); \quad \text{for} \quad x_p \in \{1, 2, ...\}. \tag{7}$$

Substituting Eq. (7) in Eq. (1), it is evident that digital projection working at the spatial Nyquist frequency produces standing fringes with fluctuating-contrast instead of phase-shifted fringes with constant-amplitude. This is illustrated in Fig. 2.

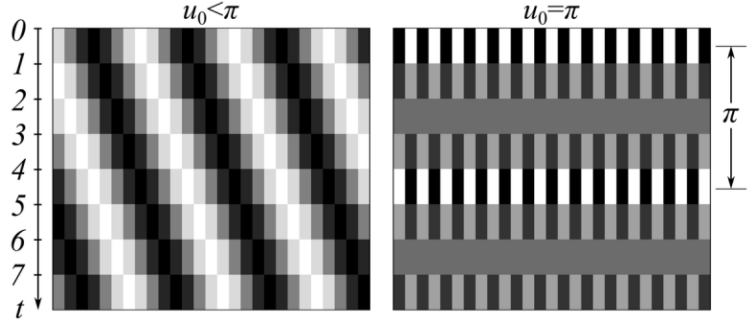

Fig. 2. Zoomed-in segment of computer-generated open fringes with spatial and temporal carriers. For the first column $u_0 < \pi$ (below Nyquist frequency); for the second column $u_0 = \pi$ (exactly at Nyquist frequency). In both cases $\omega_0 = 2\pi/8$ for illustrative purposes.

From Eq. (7) and Fig. 2, it is clear that the only linearly-independent fringe patterns with the highest-possible spatial frequency and contrast are obtained when $\omega_0 t = \{0, \pi\}$. Since the alternative is working with low-contrast fringes containing the same information and the fringes' contrast is essential for a high-quality phase-demodulation [4,5], the logical choice is to project only two fringe patterns using both Nyquist frequencies, from where we have:

$$I(x, y, t) = a(x, y) + b(x, y) \cos[\varphi(x, y) + \alpha \pi x + \pi t], \quad t = 1, 2. \quad (u_0 = \pi, \omega_0 = 0) \tag{8}$$

In contrast to conventional PSAs (having $\omega_0 < \pi$), temporal quadrature filtering at Nyquist frequency would fail to demodulate Eq. (8) because $H(-\pi) = 0$ implies $H(\pi) = 0$ due to the $2\pi$ − periodicity of the discrete-time Fourier transform [4]. This is illustrated in Fig. 3.

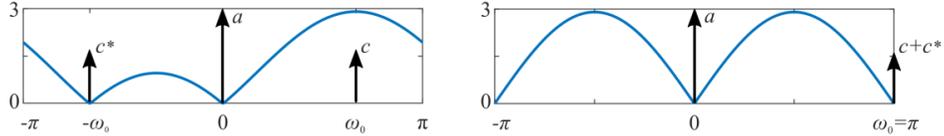

Fig. 3. Temporal quadrature filtering below and exactly at the Nyquist frequency. The vertical arrows represent the discrete-time Fourier transform of $I(x,y,t)$. The blue line is $|H_3(\omega)|$ for tunable a 3-step PSA having spectral zeroes at $\omega = \{0, -\omega_0\}$. Below Nyquist we are able to isolate $c = (b/2)\exp(i\varphi)$. On the other hand, when working at the temporal Nyquist frequency, the spectral zero at $\omega = -\omega_0$ would filter-out both analytic signals simultaneously.

Recall that defocused projection is assumed for the high-frequency binary fringes in order to obtain the sinusoidal profile modeled in Eq. (8). For a detailed discussion, see [7].

*2.2 Proposed phase-demodulation algorithm*

Here we deduce a phase-demodulation algorithm suitable for the spatial and temporal Nyquist frequencies in FPP. Operating on the phase-shifted fringe patterns on Eq. (8), we have:

$$\begin{aligned} I(x, y, 0) &= a(x, y) + b(x, y) \cos[\varphi(x, y) + \alpha \pi x], \\ I(x, y, 1) &= a(x, y) + b(x, y) \cos[\varphi(x, y) + \alpha \pi x + \pi], \\ I'(x, y) &= I(x, y, 0) - I(x, y, 1) = 2b(x, y) \cos[\varphi(x, y) + \alpha \pi x]. \end{aligned} \tag{9}$$

The subtraction in $I'(x,y)$ describes the following impulse response and FTF:

$$h_2(t) = \delta(t) - \delta(t-1), \quad H_2(\omega) = 1 - \exp(-i\omega). \qquad (10)$$

Therefore, in the temporal Fourier domain $H_2(0) = 0$ filtered the background signal while $H_2(\pi) = 2$ amplified both components at the temporal Nyquist frequency. Next, we take the spatial Fourier transform of $I'(x,y)$: $(u,v) \in (-\pi, \pi] \times (-\pi, \pi]$,

$$I'(u,v) = 2C(u - \alpha\pi, v) + 2C^*(u + \alpha\pi, v); \quad C(u,v) = F\{c(x,y)\}. \qquad (11)$$

Since $0 < \alpha < 1$, we can filter-out $C^*(u + \alpha\pi, v)$ applying a one-dimensional spatial Hilbert filter (OSHF) for the horizontal frequencies, which is defined by [10,11]:

$$H(u,v) = \begin{cases} 1 & \text{if } u > 0, \\ 0 & \text{otherwise.} \end{cases} \qquad (12)$$

Finally, back on the spatial domain, the searched the analytic signal is given by:

$$r_3(x,y)\exp[i\varphi_3(x,y)] = e^{-i\alpha\pi x} \cdot OSHF[I(x,y,0) - I(x,y,1)]. \quad (u_0 = \pi, \omega_0 = \pi) \qquad (13)$$

Here $r_3(x,y) = 2b(x,y)$. Please note that spatial low-pass filtering was not required for the phase demodulation, and that Hilbert filtering preserves all the spectral information for real-valued input data [11]. In other words, this approach is suitable to work with discontinuous surfaces, similarly to the temporal synchronous demodulation method.

## 3. Figures of merit of the proposed algorithm

### 3.1 Signal-to-noise ratio (SNR)

The analysis in the previous section was restricted to ideal conditions for ease of exposition. Now let's assume the phase-modulated fringe patterns in Eq. (3) are distorted by a zero-mean additive white-Gaussian noise (AWGN) having a flat power spectral density:

$$I_n(x,y,t) = I(x,y,t) + n(x,y,t), \qquad (14)$$

Applying Eq. (13) to $I_n(x,y,t)$, we obtain the following analytic signal:

$$r_N(x,y)\exp[i\varphi(x,y) + iN(x,y)] = r_3(x,y)\exp[i\varphi(x,y)] + OSHF[n(x,y,t) * h_2(t)], \qquad (15)$$

where $r_N \approx r_3$ for low-energy noise (a reasonable assumption for FPP), and variance of the noisy phase is $\sigma_N^2 \approx \sigma_n^2/(Mb^2)$. These results were derived for $M$-step LS-PSAs, which are defined univocally by their frequency transfer function [4,12]:

$$H_M(\omega) = \prod_{m=0}^{M-2}\{1 - \exp[-i(\omega + m\omega_0)]\}, \quad \omega_0 = \frac{2\pi}{M}. \qquad (16)$$

Comparing Eqs. (10) and (16) is obvious that our phase-demodulation method uses the 2-step LS-PSA. Typically $M$-step LS-PSAs are restricted to $M \geq 3$ in order to fulfill all quadrature conditions (see Eq. (6)). In our case, $H_2(\omega)$ satisfies $H_2(0) = 0$ but not $H_2(-\omega_0) = 0$; that condition was replaced by the one-dimensional Hilbert filtering.

Continuing the analysis of the general case for M-step LS-PSAs filtering, the signal-to-noise ratio (SNR), computed over an area $\Omega$ where we have valid data, is given by:

$$SNR_M = \frac{(1/\Omega)\int_\Omega \varphi^2(x,y)d\Omega}{(1/\Omega)\int_\Omega N^2(x,y)d\Omega} = \frac{(u_0\alpha\tan\theta)^2 \int_\Omega s^2(x,y)d\Omega}{(1/M)\int_\Omega n^2(x,y)b^{-2}(x,y)d\Omega}$$
$$= \left(\frac{M}{\Lambda^2}\right) \times \frac{(2\pi\alpha\tan\theta)^2 \int_\Omega s^2(x,y)d\Omega}{\int_\Omega n^2(x,y)b^{-2}(x,y)d\Omega}. \tag{17}$$

Here $M$ is the number of phase-shifted temporal samples, and $\Lambda$ is the number of pixels on a spatial period of the projected fringes. Note that the ratio $M/\Lambda^2$ is the only free-parameter that allows us to increase the SNR (an excessive increase of $\theta$ would produce large self-occluding shadows where $b(x,y) \to 0$). Moreover, using $\Lambda = M$ is recommendable for short spatial periods in order to avoid changes on the profile of the projected fringes. See Fig. 4.

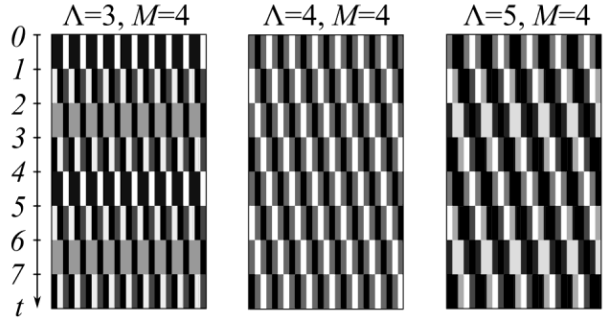

Fig. 4. Zoomed-in segment of computer-generated open fringes with spatial carrier $u_0 = 2\pi/\Lambda$ and temporal carrier $\omega_0 = 2\pi/M$. Note that only $\Lambda = M$ produces a non-varying fringes' profile.

Comparing the SNR for our proposed method (where $\Lambda = M = 2$) against other approaches using $\Lambda = M$ with $M \geq 3$, we found our 2-step method to have the highest SNR:

$$\frac{SNR_2}{SNR_M} = \frac{(1/2)}{(1/M)} = \frac{M}{2} > 1 \quad \forall M \geq 3 \quad (\text{for } \Lambda = M). \tag{18}$$

Equation (18) assumes high-frequency spatial carriers. The variations on the fringes' profile are less significant for low-frequency spatial carriers, leaving $M/\Lambda^2$ unrestricted. As usual, the SNR for this case is improved using as many phase-shifted temporal samples as possible.

*3.2 Robustness against distorting harmonics*

Harmonic distortion is to be expected under saturation on the image sensor and/or due to residual distortion after the defocused projection of the binary fringes [1-5,7,8]. Therefore, a more realistic model for our phase-modulated fringe patterns is given by:

$$I(x,y,t) = \sum_{k=-K}^{K} b_k \exp[ik(\varphi + \alpha\pi x + \pi t)]; \quad (x,y) \in \mathbb{Z}^2, \quad t = 0,1. \tag{19}$$

Here $b_k$ is the contrast of the *k*-th order distorting harmonic, satisfying $b_{-k} = b_k \; \forall k$, and $|b_{k+1}| < |b_k|$ in general. $K$ is the order of the distorting harmonic with still-significant energy. The background signal or DC-offset is given by $b_0$, and all other terms remains as previously defined. As before, the ideal result would mean to isolate $c_1 = b_1 \exp(i\varphi)$.

Substituting Eq. (19) in Eq. (9), we have after some algebraic manipulation:

$$I'(x,y) = I(x,y,0) - I(x,y,1) = \sum_{k=-K}^{K}[1-(-1)^k]b_k \exp[ik(\varphi+\alpha\pi x)]. \qquad (20)$$

This result shows that the 2-step temporal filter eliminates not only the background signal ($k=0$) but also filters-out all distorting harmonics of even order. For completeness (we will revisit this result at the end of this section), we have in the discrete-time Fourier domain:

$$H_2(\omega)I(x,y,\omega) = (1-e^{-i\omega})\sum_{k=-K}^{K} b_k e^{ik(\varphi+\alpha\pi x)}\delta(\omega-\omega_k); \quad \omega_k = \mathrm{Arg}(e^{i\pi k}). \qquad (21)$$

Note that some even-order harmonics are aliased to the spectral origin and filtered-out by $H_2(0)=0$, while the analytic signals $\{b_1 e^{i\varphi}, b_1 e^{-i\varphi}\}$ and every odd-order harmonic are aliased to the spectral frequency $\omega = \pi$, where they are amplified by $H_2(\pi) = 2$.

Back to Eq. (20), taking the discrete 2D Fourier transform we have:

$$I'(u,v) = \sum_{k=-K}^{K}[1-(-1)^k]C_k(u-u_k,v); \quad C_k(u,v) = F\{b_k e^{ik\varphi}\}; \quad u_k = \mathrm{Arg}(e^{i\pi\alpha k}). \qquad (22)$$

The Hilbert filtering at Eq. (13) erases any spectral component within the negative-half of the spatial spectrum, but only $F\{b_1 \exp(-i\varphi)\}$ is guaranteed to be there: $u_{-1} = -\alpha\pi$. For $k \geq 3$, one component of every $k$-th odd-order harmonic lies within the positive-half of the spectral domain, where $0 > u > \pi$, and they remain unfiltered by the OSHF. Therefore, we conclude from Eqs. (13)-(22) that the analytic signal distorted by harmonics will be given by:

$$r_3(x,y)\exp[i\varphi_3(x,y)] = b_1 e^{i\varphi} + b_3 e^{\pm i3\varphi} + b_5 e^{\pm i5\varphi} + b_7 e^{\pm i7\varphi} + \dots \qquad (23)$$

The plus-or-minus signs depend on the actual value of $\alpha$. As shown in Fig. 5, the robustness against harmonics of our 2-step phase-demodulation algorithm is superior to the 3-step LS-PSA (particularly because the second harmonic is usually the one with highest distorting energy) and equivalent the robustness of the popular 4-step LS-PSA.

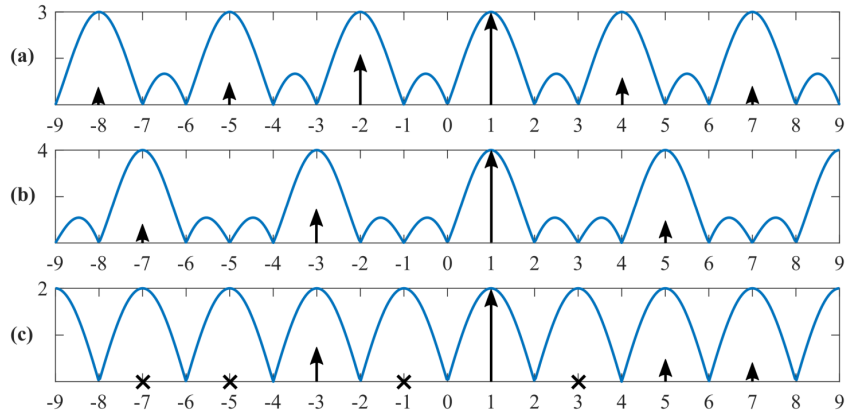

Fig. 5. Harmonics-rejection capabilities of the (a) 3-step and (b) 4-step LS-PSAs versus (c) our 2-step algorithm for $1/4 < \alpha < 1/2$ (without loss of generality). The crosses in the third plot represent harmonics rejected by the Hilbert filtering process. The horizontal axes correspond to normalized frequency ($\omega/\omega_0$) beyond the wrapped range for ease of observation [4].

## 4. Experimental results

Feasibility of the proposed method was confirmed with experimental results using a 1-camera 1-projector profilometer (as shown in Fig. 1). We used a digital light processing (DLP) projector with $1024\times768$ pixels resolution and 8-bit color depth per channel, working only with the green channel for monochromatic operation. The images were recorded using an 8-bit grayscale digital camera with $1600\times1200$ pixels resolution. The camera explored a small fraction of the region being illuminated by the DLP projector. Calibrating the effective carrier with help of the reference plane we found $\alpha = 0.2167$ for our particular setup.

Two 3D solids were digitized: a continuous surface as calibration object (showing the many-wavelengths sensitivity achieved) and a piece-wise continuous object (illustrating the applicability of the proposed method for the study of discontinuous surfaces).

*4.1 Digitization of a continuous surface*

Figures 6-8 show the main stages of the digitization of a smooth surface: a metallic spherical cap. The high-frequency open fringes are difficult to see at this small scale but satisfactory contrast was achieved on the raw data (see Fig. 6(b)). In Fig. 7 intensity is in logarithmic scale with the spectral origin at the center for ease of visualization. Note from Fig. 8 the many-wavelength sensitivity achieved projecting at the spatial Nyquist frequency.

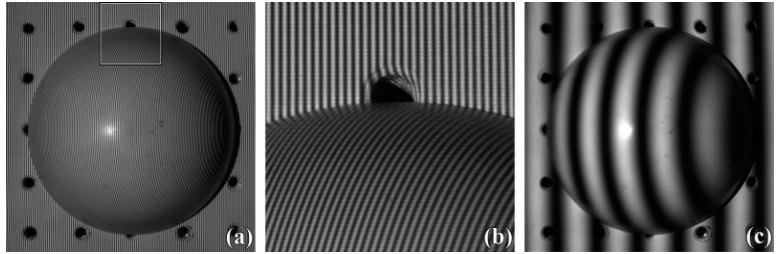

Fig. 6. Fringe patterns phase-modulated by a spherical cap. (a) One of the two fringe patters obtained with fringe projection using both Nyquist frequencies, $u_0=\pi$ and $\omega_0=\pi$. (b) Zoomed-in section of (a). (c) Low-frequency carrier fringe pattern using $u_0=\pi/20$, for visual comparison.

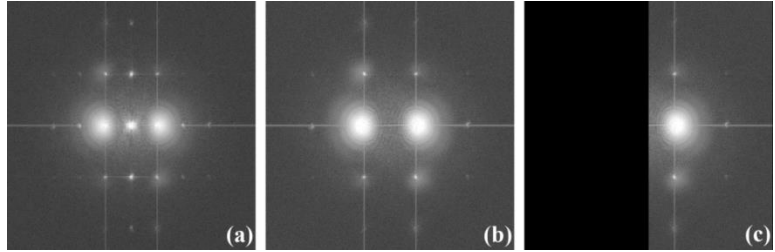

Fig. 7. Spatial Fourier spectra of the fringe patterns phase-modulated by the spherical cap. (a) corresponds to the raw data, (b) is obtained after temporal filtering, and (c) represents the Hilbert filter that rejects the negative-frequency spectral components. Note that the spectral lobes which correspond to the DC component and odd-order harmonics are gone from (b).

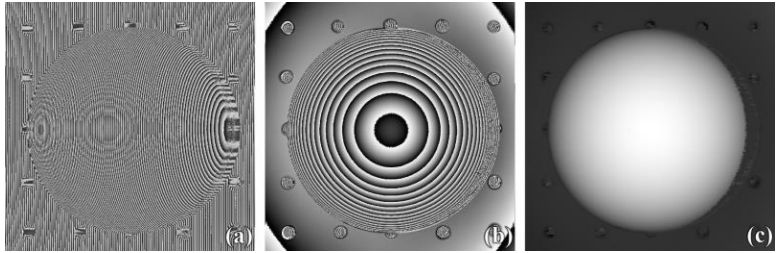

Fig. 8. Phase maps computed from the analytic signal in Eq. (13). For illustrative purposes, (a) shows the highly wrapped phase before subtracting the spatial carrier. (b) Shows the highly wrapped phase after subtracting the spatial carrier. (c) Shows the phase map proportional to the object's height profile, computed with spatial unwrapping from (b).

*4.2 Digitization of a discontinuous surface*

Figures 9-11 show the main steps on the digitization of a discontinuous (piece-wise continuous) solid: a combination-square tool equipped with the standard head. As before, we worked at the Nyquist frequencies to obtain a high-sensitivity highly wrapped phase map as the argument of Eq. (13). However, the surface steps and the self-occluding shadows made impossible to unwrap the highly modulated phase-map using spatial methods. Figure 9(c) shows one of four low spatial-frequency fringe patterns processed additionally to obtain a non-wrapped low-sensitivity estimation (having $u_0 = \pi/10$ and $\omega_0 = 2\pi/4$), which was used as stepping stone for dual-sensitivity temporal phase unwrapping of Fig. 11(b) [13].

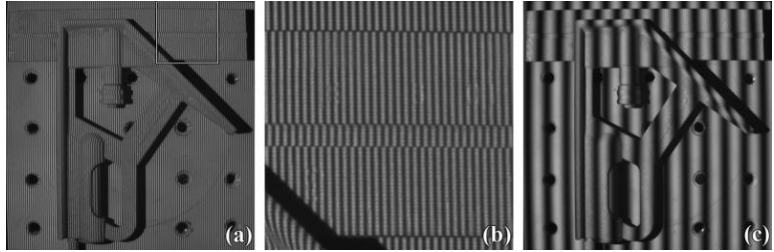

Fig. 9. Digitization of a combination-square tool. (a) One of two phase-modulated fringe patterns obtained with digital fringe projection using both Nyquist frequencies. (b) Zoomed-in section of (a). (c) One of four low-frequency carrier fringes (using $u_0=\pi/10$) to be used as stepping stone on dual-sensitivity temporal phase unwrapping [13].

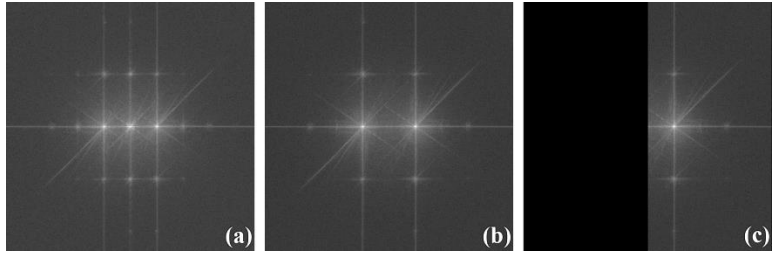

Fig. 10. Spatial Fourier spectra of the fringe patterns modulated by the combination square. First panel corresponds to the raw data, (b) is obtained after temporal filtering, and (c) represents the Hilbert filter rejecting the negative-frequency spectral components.

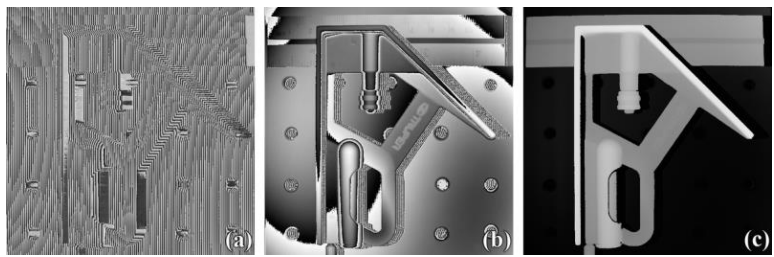

Fig. 11. Phase maps computed from the searched analytic signal. (a) Highly wrapped phase modulated by the object and the spatial carrier. (b) Wrapped phase after subtracting the spatial carrier. (c) Unwrapped phase map proportional to the searched height profile, computed from (b) and dual-sensitivity temporal phase unwrapping [13].

## 5. Conclusions and perspectives

We proposed a robust and easy to implement phase-demodulation method for digital FPP suitable for the spatial and temporal Nyquist frequencies during the projection stage. We remark, once again, that the Nyquist-criterion must be satisfied during the recording stage (having $\alpha < 1$ in our notation). Using the spatial Nyquist frequency ensures the highest depth-sensitivity for any given angle between the projection and recording directions. The proposed method does not require any low-pass filter for the phase-demodulation, so it can be applied to the digitization of discontinuous (piece-wise continuous) surfaces.

Among the many phase-demodulation algorithms based on DC bias removal plus Hilbert transform/filtering, the only one we found to be equivalent to Eq. (13) was published by Zheng and Da in 2013 [14]. However, its applicability to the Nyquist frequencies was not even mentioned. Instead they discussed a method for gamma compensation, because high-order distorting harmonics (due to non-linear gamma encoding) usually require many-step PSAs for high-quality estimations [1-5]. Nevertheless, we demonstrated in this paper that the proposed algorithm (Eq. (13)) is as robust against distorting harmonics as the popular 4-step LS-PSA despite using half the number of phase-shifted temporal samples. Moreover, non-linear gamma calibration is unnecessary for our method because the projected fringes become binary at the spatial Nyquist frequency.

The proposed method is convenient for high-speed implementations because it requires a single phase-step (two frames) and, as illustrated in Fig. 2, the on/off pixels alternate between each phase-shifted fringe pattern. This makes them ideal for color-encoded single-image FPP. For instance, one might encode the first frame in the red channel and the one shifted by $\pi$ radians (equivalently to a 1-pixel lateral shift) in the blue channel. The resulting full-color image will contain both phase-shifted fringes patterns with essentially null-crosstalk and without spatial overlap (the later meaning the 3D test surface will be fully illuminated the entire time, except of course for self-occluding shadows).

### Acknowledgments

The authors want to thank Cornell University for supporting the e-print repository arXiv.org and the Optical Society of America for allow posting the contributed manuscripts at arXiv.